\documentclass[11pt]{article}
\usepackage[a4paper,margin=1in]{geometry}
\usepackage{amsmath,amssymb,amsthm,bm,mathtools}
\usepackage{graphicx}
\usepackage{microtype}
\usepackage{booktabs}
\usepackage{xcolor}
\usepackage{setspace}
\usepackage[super,sort&compress]{natbib}
\usepackage[hidelinks]{hyperref}
\IfFileExists{newtxtext.sty}{\usepackage{newtxtext,newtxmath}}{}
\newcommand{\STF}{\operatorname{STF}}
\newcommand{\Vtx}{\mathcal V}
\newcommand{\cT}{\mathcal T}
\newcommand{\dd}{\mathrm d}
\newcommand{\e}{\mathrm e}
\newcommand{\R}{\mathbb R}
\newcommand{\Harm}{\mathcal H}
\newcommand{\Proj}{\Pi_0}

\title{One vertex generates the hard-sphere Boltzmann collision hierarchy}
\author{Ilya Karlin$^{1}$}
\date{October 2026}

\begin{document}
\maketitle
\begin{center}
\small $^{1}$Department of Mechanical and Process Engineering, ETH Zurich, CH-8092 Zurich, Switzerland\\
Correspondence to: ikarlin@ethz.ch
\end{center}
%\linenumbers

\noindent{The Boltzmann equation is the fundamental kinetic bridge between molecular mechanics and fluid behaviour. It converts binary encounters into the evolution of a velocity distribution, underlies rigorous routes from hard-sphere dynamics to Euler and Navier--Stokes--Fourier equations, and remains indispensable when continuum descriptions fail, from hypersonic shock layers and atmospheric re-entry to rarefied multiscale flows.\cite{Boltzmann1872,DengHaniMa2024,DengHaniMa2025,ReentrySurvey2020,GuoXu2021} Yet its nonlinear collision term is conventionally represented by an infinite three-index tensor of collisional brackets, with apparently new coefficients at every polynomial order. Here we prove that this entire hard-sphere tensor contains only one primitive collision law: the exact rank-two stress-production vertex $S$. Every polynomial collision vertex is obtained from $S$ by a finite sequence of universal differential and tensor operations, and every conventional bracket follows by differentiation. Thus the interaction physics and its representation separate exactly: all hard-sphere-specific information is concentrated in $S$, whereas the hierarchy above it is universal algebra. A blind eighth-order test in a Mach-5 shock agrees with independent collision sampling and strongly amplifies a weak kinetic precursor. 
The hard-sphere hierarchy is infinite in dimension, but its primitive collision content consists of a single scalar law.
}

\newpage

The Boltzmann equation is the fundamental kinetic equation for a dilute gas. It occupies a unique position between Newtonian particle dynamics and continuum fluid mechanics: microscopic mechanics enters through the collision law, while velocity moments of the distribution generate density, momentum, energy and, in hydrodynamic limits, the Euler and Navier--Stokes--Fourier equations. This molecular-to-fluid passage was already central to Hilbert's sixth problem, and recent work has carried the hard-sphere derivation from particle dynamics through Boltzmann kinetics to fluid equations over regimes far beyond the classical short-time theory.\cite{Hilbert1902,DengHaniMa2024,DengHaniMa2025} In applications, the same equation becomes unavoidable precisely where the continuum approximation is least reliable, including high-altitude hypersonics, atmospheric re-entry, rarefied spacecraft aerodynamics and multiscale gas flows.\cite{ReentrySurvey2020,GuoXu2021}

For identical hard spheres, we use the dimensionless collision operator
\begin{equation}
Q_{\rm HS}(f,f)(v)=\sqrt\pi
\int_{\R^3}\!\dd v_*\int_{S^2}\!\frac{\dd\sigma}{4\pi}\,
|v-v_*|\,[f(v')f(v_*')-f(v)f(v_*)],
\label{eq:QHS}
\end{equation}
where $C=(v+v_*)/2$, $g=|v-v_*|$, $v'=C+(g/2)\sigma$ and $v_*'=C-(g/2)\sigma$. Equation~\eqref{eq:QHS} is compact. Its conventional moment representation is not.

Expand the distribution around a Gaussian reference $M$ in an orthonormal polynomial basis, $f=M\sum_\lambda a_\lambda\Psi_\lambda$. The production of every outgoing coefficient has the quadratic form
\begin{equation}
\dot a_\lambda=\sum_{\mu,\nu}P_{\lambda\mu\nu}\,a_\mu a_\nu,
\qquad
P_{\lambda\mu\nu}=\frac{\sqrt\pi}{2}\iiint MM_*\,g\,\Psi_\mu(v)\Psi_\nu(v_*)\,\Delta\Psi_\lambda\,
\frac{\dd\sigma}{4\pi}\dd v_*\dd v ,
\label{eq:Ptable}
\end{equation}
where $\Delta\Psi_\lambda=\Psi_\lambda(v')+\Psi_\lambda(v_*')-\Psi_\lambda(v)-\Psi_\lambda(v_*)$. The coefficients $P_{\lambda\mu\nu}$ are the nonlinear collisional brackets. In Cartesian Hermite coordinates the labels are multi-indices; in an irreducible Burnett basis they resolve into radial and angular quantum numbers. Either way, increasing the polynomial degree exposes an expanding three-index table of exact hard-sphere collision data. Classical Hermite, Burnett and Sonine constructions organize this table efficiently, and rotational symmetry eliminates many entries, but the table itself remains the conventional object to be computed.\cite{Grad1949,Kumar1966}

\begin{figure}[t]
\centering
\includegraphics[width=0.98\linewidth]
{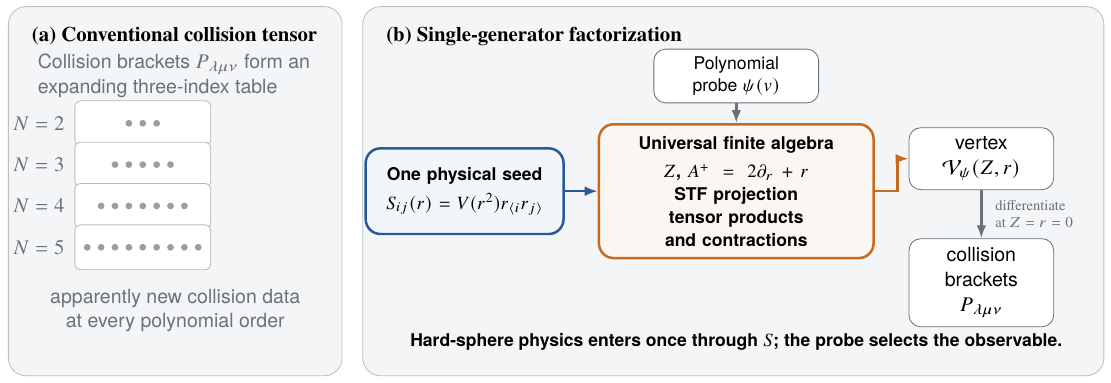}
%{fig1_classification_streamlined.pdf}
\caption{\textbf{Classification of the nonlinear hard-sphere collision tensor.} \textbf{a}, A conventional polynomial representation exposes successive slices of the three-index collisional-bracket tensor $P_{\lambda\mu\nu}$, with apparently new collision data at every degree. \textbf{b}, The construction has two inputs: the stress-production vertex $S$ supplies the complete interaction-specific content, while the chosen probe $\psi$ selects which finite combination of universal operations is applied to it. The resulting vertex $\mathcal V_\psi$ generates every conventional bracket in that outgoing row by differentiation.}
\label{fig:classification}
\end{figure}

The structural question is therefore more basic than how to evaluate a particular bracket: \emph{how many independent pieces of hard-sphere collision information are actually contained in the tensor $P_{\lambda\mu\nu}$?} Figure~\ref{fig:classification} summarizes the answer.

\section*{A single hard-sphere generator}

Introduce coherent labels $z$ and $u$ on the two incoming Hermite legs and the combinations
\begin{equation}
Z=z+u,\qquad r=z-u,\qquad s=r^2.
\label{eq:Zr}
\end{equation}
For each outgoing polynomial observable $\Psi_\lambda$, its coherent collision vertex $\Vtx_\lambda(z,u)$ is the exponential generating function of the conventional brackets,
\begin{equation}
\Vtx_\lambda(z,u)=\sum_{\mu,\nu}P_{\lambda\mu\nu}\,
\frac{z^\mu u^\nu}{\sqrt{\mu!\nu!}},
\qquad
P_{\lambda\mu\nu}=
\left.\frac{\partial_z^\mu\partial_u^\nu\Vtx_\lambda(z,u)}{\sqrt{\mu!\nu!}}\right|_{z=u=0}.
\label{eq:bridge}
\end{equation}
Multi-index derivatives in the coherent labels span both incoming Hermite legs, so Eq.~\eqref{eq:bridge} is exhaustive: every polynomial bracket is a vertex Taylor coefficient, and conversely.

The primitive object is the exact stress-production vertex,
\begin{equation}
S_{ij}(r)=V(s)\,r_{\langle i}r_{j\rangle},
\label{eq:Svertex}
\end{equation}
where angular brackets denote the symmetric traceless part and $V$ is the scalar collisional law \cite{Karlin2026vertex}:
\begin{equation}
{
V(s)=-\frac12\frac{\e^{-s/4}(s^2+4s-12)}{s^2}
-\frac{\sqrt\pi}{4}\frac{s^3+6s^2-12s+24}{s^{5/2}}
\operatorname{erf}\!\left(\frac{\sqrt s}{2}\right).}
\label{eq:Vexact}
\end{equation}
The apparent singularity at $s=0$ is removable; the entire continuation begins $V(s)=-8/5-(2/35)s+(1/1260)s^2+\cdots$. Equation~\eqref{eq:Vexact} is the only interaction-specific scalar law required by the complete polynomial hierarchy. The conventional heat-flux production vertex is already its first mixed descendant \cite{Karlin2026vertex},
\begin{equation}
\Vtx^{\rm heat}_i(Z,r)=\frac12 Z_jS_{ij}(r),
\label{eq:heat}
\end{equation}
but the same principle continues at every polynomial order.

The content of the theorem is most transparent for the monic symmetric
trace-free (STF) probes commonly used in kinetic theory,
\begin{equation}
 \psi^{(N)}(v)=v_{\langle i_1}\cdots v_{i_N\rangle}.
\label{eq:monicprobe}
\end{equation}
Let $A_i^+=2\partial_{r_i}+r_i$ be the relative Gaussian raising operator.  Then
the complete vertex selected by the degree-$N$ probe is
\begin{equation}
\boxed{
 \Vtx^{(N)}(Z,r)
 =2^{2-N}\sum_{q=1}^{\lfloor N/2\rfloor}
 \binom N{2q}
 \STF\!\left[
 Z^{\otimes(N-2q)}\otimes
 (A^+)^{\otimes(2q-2)}\otimes S(r)
 \right].}
\label{eq:monicgeneration}
\end{equation}
Here the STF projection acts on all
free indices. Equation \eqref{eq:monicgeneration} demonstrates that the vertex collapses to the primitive $S$ \eqref{eq:Svertex} through a finite string of universal algebraic operations. More generally, this collapse happens to any polynomial probe:

\noindent\textbf{Single-generator theorem.} \emph{On the Hermite polynomial
core of the three-dimensional hard-sphere Boltzmann collision map, the
stress-production vertex $S$ generates every non-invariant outgoing collision
vertex.  Equation~\eqref{eq:monicgeneration} gives the closed monic sequence.
An arbitrary scalar-, vector- or tensor-valued polynomial probe is reduced by a
finite Wick--Fischer decomposition to the same descendants of $S$, using only
multiplication by $Z$, the commuting operators $A_i^+$, their scalar contraction
$A^+\!\cdot A^+$, tensor contraction and irreducible $SO(3)$ projection.  Hence
every conventional Hermite or Burnett bracket is obtained by finite
differentiation, and no second hard-sphere collision function appears.}

The fully general formula, with the finite coefficients read directly from the
pair polynomial of the chosen probe, is given and proved in Supplementary
Sections~2--7.  Keeping that bookkeeping in the Supplement leaves the physical
claim visible here: vertex $=$ universal algebra applied to $S$.

\section*{Why the collapse occurs}

The proof is short once the collision is written in centre and relative variables. For any polynomial observable $\psi(v)$, define the symmetrized pair polynomial
\begin{equation}
\Phi_\psi(C,G)=\psi(C+G/2)+\psi(C-G/2),\qquad G=v-v_*.
\label{eq:pairsym}
\end{equation}
It is even in $G$. At fixed $C$ and $g=|G|$, hard-sphere scattering in Eq.~\eqref{eq:QHS} replaces the incoming direction of $G$ by an isotropically distributed outgoing direction. The angular operator entering the pair weak form is therefore
\begin{equation}
\mathcal L_g^{\rm HS}=\frac{\sqrt\pi}{2}g(\Proj-I),
\label{eq:angular}
\end{equation}
with $\Proj$ the spherical average.

To make this key observation explicit, resolve the $G$-dependence into irreducible
$SO(3)$ components labelled by the rank $L$ of the solid harmonic
$\Harm_L(G)=G_{\langle i_1}\cdots G_{i_L\rangle}$.  Since $\Phi_\psi$ is even
in $G$, only $L=0,2,4,\ldots$ occur.  The $L=0$ component is precisely the
spherical average $\Proj\Phi_\psi$; Eq.~\eqref{eq:angular} removes it and
assigns the same non-zero hard-sphere rate to every remaining even sector
$L>0$.  This complete degeneracy of the active angular spectrum is the
interaction-specific reason that a single generator is possible.

Fischer decomposition now resolves the active part of any even relative
polynomial into the finite sum
\begin{equation}
(1-\Proj)\Phi_\psi
=\sum_{p\ge0}\sum_{L=2,4,\ldots}A_{pL}(C):[(G^2)^p\Harm_L(G)],
\label{eq:Fischer}
\end{equation}
with finite coefficient tensors $A_{pL}(C)$. 
Every even solid harmonic is an STF product of the rank-two harmonic
$\Harm_2(G)$.  To pass these relative polynomials through the remaining
Gaussian integral, define the hard-sphere weighted transform
\begin{equation}
 \cT[P](r):=\frac{\sqrt\pi}{2}\int_{\R^3}\rho_r(G)\,|G|\,P(G)\,\dd G,
 \qquad
 \rho_r(G)=(4\pi)^{-3/2}\exp\!\left[-\frac{|G-r|^2}{4}\right].
\label{eq:Tdef}
\end{equation}
For real $r$, $\rho_r$ is the density of normally distributed random vector $G\sim N(r,2I)$; for complex $r$ the
quadratic form in the exponent is interpreted bilinearly and the coherent
vertex is obtained by entire continuation.  Direct differentiation gives the exact
insertion rule $\cT[G_iP]=A_i^+\cT[P]$, with
$A_i^+=2\partial_{r_i}+r_i$.  Hence
\begin{equation}
\cT[(G^2)^p\Harm_L]
=(A^+\!\cdot A^+)^p
\STF[(A^+)^{\otimes(L-2)}\otimes\cT[\Harm_2]].
\label{eq:intertwine}
\end{equation}
The identity $\cT[\Harm_2]=-2S$ then identifies the transformed rank-two harmonic with the primitive vertex, while centre Wick tensors average to ordinary polynomials in $Z$. Pair symmetry, the hard-sphere angular collapse and the Gaussian intertwiner therefore generate every surviving term from this seed. The complete proof, including the tensor-valued Fischer decomposition, is given in Supplementary Sections 2--6.

A first view of the generated hierarchy follows directly from
Eq.~\eqref{eq:monicgeneration}:
\begin{align}
 \Vtx^{(2)}&=S,\nonumber\\
 \Vtx^{(3)}&=\frac32\STF[Z\otimes S],\nonumber\\
 \Vtx^{(4)}&=\frac32\STF[Z^{\otimes2}\otimes S]
 +\frac14\STF[(A^+)^{\otimes2}\otimes S].
\label{eq:firstdescendants}
\end{align}
Thus the first genuinely new angular row already contains no new scalar
collision function. Its linearization reproduces independently computed
hard-sphere rank-four matrix elements, including radial descendants
(Supplementary Section~9).

\section*{Physics separated from representation}

The theorem separates two ingredients that are intertwined in a conventional
collision table.  The first is physical: the relative-speed factor and
hard-sphere scattering law determine the single scalar function $V(s)$ \eqref{eq:Vexact} and
hence the tensor $S$.  The second is representational: once $S$ is known,
raising by $A^+$, multiplication by $Z$, tensor contraction and $SO(3)$ projection place
that physical content into whichever polynomial moment is being observed.
These latter operations are universal.  They neither use the collision cross
section again nor create a new interaction-dependent scalar law.

The familiar collisional brackets are recovered immediately after this
construction.  Since $z=(Z+r)/2$ and $u=(Z-r)/2$, it follows 
%\begin{equation}
$\partial_{z_i}=\partial_{Z_i}+\partial_{r_i}$,
$\partial_{u_i}=\partial_{Z_i}-\partial_{r_i}$.
%\end{equation}
For a normalized Cartesian Grad--Hermite probe
$\Psi_\lambda(v)=\operatorname{He}_\lambda(v)/\sqrt{\lambda!}$, the complete
conventional row is therefore
\begin{equation}
{
 P_{\lambda\mu\nu}
 =\left.
 \frac{(\partial_Z+\partial_r)^\mu
       (\partial_Z-\partial_r)^\nu}
      {\sqrt{\mu!\nu!}}
 \Vtx_{\Psi_\lambda}(Z,r)
 \right|_{Z=r=0}.}
\label{eq:recoverbracketsZr}
\end{equation}
The outgoing label $\lambda$ selects the probe; differentiation exposes its two
unrestricted incoming Hermite legs.  Thus a new polynomial moment requires a
new finite arrangement of the same seed, not a new collision integral.

Cartesian Hermites, irreducible Burnett polynomials and other polynomial bases
are consequently different coordinate systems on the same generated object.
Within each fixed degree they are related by a finite basis transformation.
Such transformations may be valuable computationally, but they belong to the
representation layer and cannot introduce another primitive vertex.  This is
stronger than symmetry compression alone \cite{Hiemstra2026}: 
the collision data are not merely organized more efficiently; for hard spheres they collapse to the one non-equilibrium seed $S$.

\section*{An eighth-order kinetic spectrometer}

\begin{figure}[t]
\centering
\includegraphics[width=0.82\linewidth]{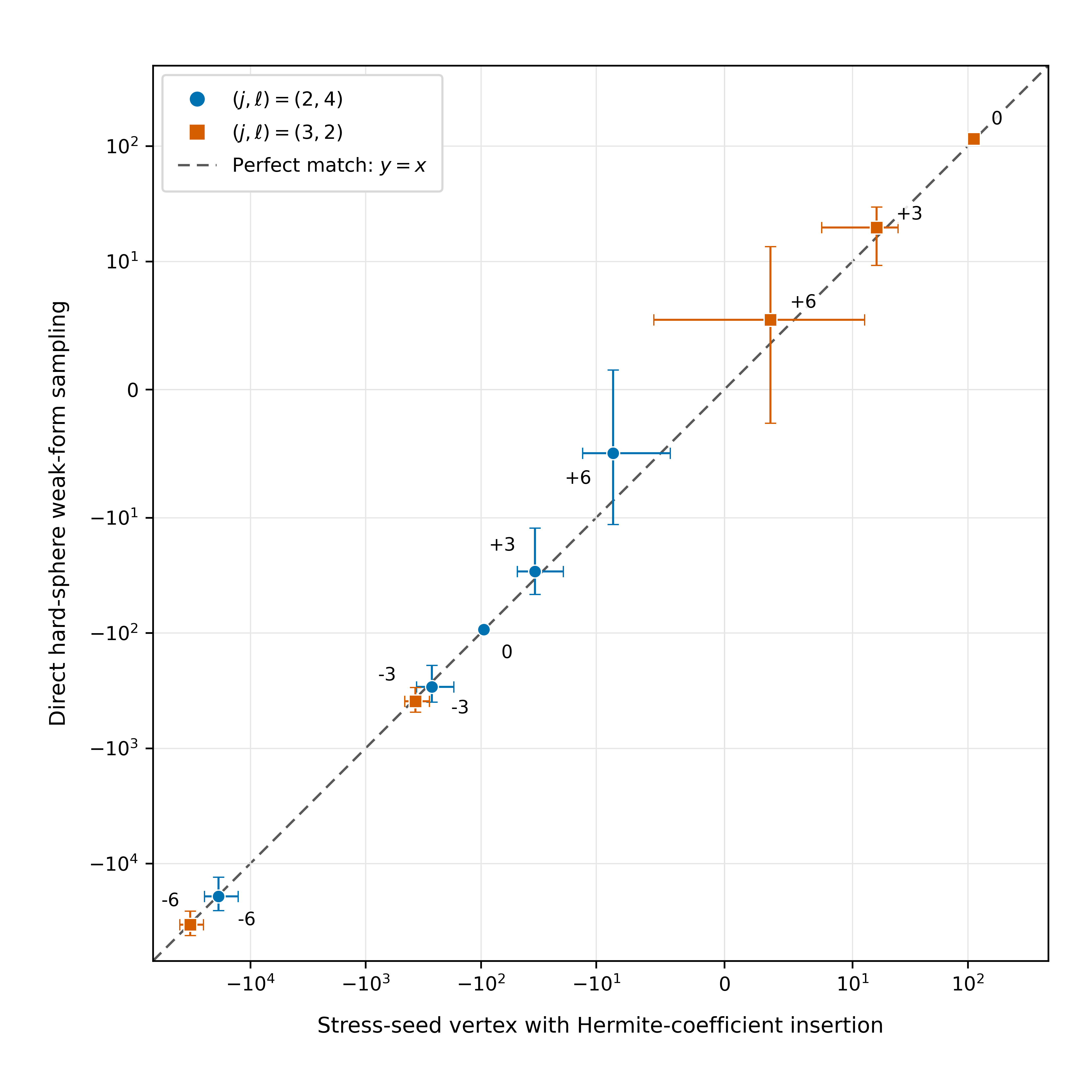}
% Replacement caption for Figure 2 in the main text.
% Insert in the existing figure environment; no regression or fitted slope.
\caption{\textbf{Degree-eight hard-sphere productions in a Mach-5 shock.}
Stress-seed predictions obtained by inserting measured Cartesian Hermite
coefficients are compared with direct sampling of the post-minus-pre collision
weak form.  Circles and squares denote the longitudinal unit-normalized Burnett
probes $(j,\ell)=(2,4)$ and $(3,2)$, respectively.  Labels give cell-width
offsets $-6,-3,0,+3,+6$ from the instantaneous shock density midpoint; the
nearest cell is sampled.  The coefficient calculation uses Gaussian reference
variance $\Theta=2T$ and all incoming modes through total degree $N=24$ on
each leg.  Both axes show the same density-weighted production in the fixed
normalization $\eta_{\mathrm{HS}}=\sqrt\pi/2$ multiplying the complete pair
increment.  The dashed diagonal is perfect agreement, $y=x$.  Points are means
over four independent DSMC replicas; horizontal and vertical bars are one
standard error of the respective means.  The two estimates share velocity
samples and their errors are correlated.  Both axes use the same symmetric
logarithmic scale, linear between $-10$ and $10$.  Supplementary Section~10 gives
the sampling protocol, paired uncertainties and incoming-cutoff study.}
\label{fig:m5}
\end{figure}

We tested the classification on two deliberately traced degree-eight observables, $(j,\ell)=(2,4)$ and $(3,2)$, in a stationary Mach-5 hard-sphere shock DSMC simulation. These rows contain several radial and angular descendants and are not top-rank continuations of stress. The horizontal coordinate in Fig.~\ref{fig:m5} is the exact single-generator theorem prediction; the vertical coordinate is an independently sampled hard-sphere weak-form production. No degree-eight coefficient was fitted.

Across ten aggregated measurements the slope through the origin is $1.00179$, the normalized root-mean-square mismatch is $0.180\%$, and the largest paired deviation is $1.30$ standard errors. Agreement therefore survives strongly non-equilibrium distributions and branches that were not used to identify the stress seed.

The dynamic range in Fig.~\ref{fig:m5} is itself physically informative. The
finite algebraic coefficients remain of order unity; the enormous upstream
signal is supplied by the distribution. A small population of particles
leaking from the hot downstream side lies many local thermal widths from the
cold, nearly Gaussian upstream core. An eighth-order observable weights that
weak population by a high power of velocity, and the hard-sphere kernel
contributes an additional relative-speed factor. The generated high-order
collision row therefore acts as a velocity-space spectrometer: it reveals a
kinetic precursor that is almost invisible in density, mean velocity and
temperature.

\section*{A different organization of kinetic theory}

The conventional hard-sphere hierarchy is infinite, but its primitive nonlinear collision content is not. Stress changes status in this classification. It is not merely the second-order transport observable familiar from continuum mechanics; its nonlinear vertex is the lowest active generator of the complete polynomial collision hierarchy. The traditional implication
\[
\text{new moment}\;\Rightarrow\;\text{new collision integral}
\]
is replaced by
\[
\text{new moment}\;\Rightarrow\;\text{finite algebraic descendant of one exact vertex}.
\]
This replacement is also a separation principle. Molecular interaction enters
once, through $S$; the passage from $S$ to stress, heat flux and arbitrarily
high polynomial moments is representation-independent algebra. A change of
moment basis changes the coordinates of the descendants, not their physical
source.

The result is special to hard spheres rather than a generic consequence of
rotational symmetry. For a homogeneous kernel $B(g,\mu)=g^\gamma b(\mu)$,
different angular sectors generally carry distinct rates
$\nu_L(g)=\beta_Lg^\gamma$. Partial collapse can nevertheless survive: stress
and heat flux occupy the same $L=2$ orbit, and homogeneous power laws share one
radial shape while retaining the discrete angular data $\{\beta_L\}$. Hard
spheres are exceptional because isotropic scattering makes every active $L>0$
sector share the same rate, so these separate orbits merge into one. Maxwell
molecules possess a different exact simplification \cite{Bobylev1975Fourier}---grading from the absence
of relative-speed dependence---but a generic Maxwell angular kernel retains
infinitely many angular collision data. Hard spheres exchange simple grading
for the stronger one-generator property.

The theorem is intentionally stated on the Hermite polynomial core. Extending the cyclic representation to a completed nonlinear collision operator requires separate domain and convergence control. That functional-analytic question is distinct from the finite exact classification proved here. At every polynomial degree the conclusion is complete: the three-index collision tensor contains infinitely many entries but only one primitive hard-sphere law.

\section*{Methods}
\paragraph{Polynomial-core proof.}
The proof is given in Supplementary Sections 2--6. Pair symmetrization makes the outgoing pair polynomial even in the relative variable $G$. Tensor-valued Fischer decomposition then produces only $(G^2)^p\Harm_L$ with even $L$. The hard-sphere angular operator acts by the same scalar on all $L>0$ sectors, and the shifted Gaussian transform converts multiplication by $G$ into the relative raising operator $A^+$. Centre Wick tensors average to polynomials in $Z$. These steps establish finite generation from $\Harm_2$ and hence from $S$.

\paragraph{Finite construction for a chosen probe.}
Given a polynomial observable, form its pair lift $\Phi_\psi(C,G)$, separate
its finite centre polynomial from its Fischer components in $G$, and remove the
relative scalar sector. Each surviving $(G^2)^p\Harm_L(G)$ is replaced by the
corresponding scalar and STF raising operations applied to $S$; centre Wick
tensors become ordinary polynomials in $Z$. This Cartesian procedure proves
and computes the result without choosing an oscillator representation. A
normalized Hermite or Burnett row is obtained afterward by a finite within-grade
basis transformation (Supplementary Sections~7 and 8).

\paragraph{Mach-5 validation.}
The stationary shock uses a monatomic gas with upstream Mach number 5, $T_1=1$, 50 cells and upstream mean free path $\lambda_1/L=0.05$. 
Four independent replicas were run, each with a target population of approximately $1.8\times10^4$ particles.
 Five positions relative to the instantaneous density midpoint were sampled. The direct estimator averaged isotropically sampled post-collision directions for random local pairs; the generated estimator used the exact collision-active Wick--Fischer projection. Statistics were accumulated in independent blocks. Full details are given in Supplementary Section 10.

\paragraph{Acknowledgement of AI assistance.}
During the development and preparation of this manuscript, the author used
ChatGPT (OpenAI, GPT-5.6 Sol) as an interactive research and writing assistant.
All results and final text were reviewed and accepted by the
author, who assumes full responsibility for the content of the manuscript.

\end{document}